\documentclass[preprint,12pt]{elsart}
\usepackage{graphicx,amssymb,amsmath,times}
\usepackage{setspace}
\journal{New Astronomy}

\begin{document}
\begin{frontmatter}
\title{Artificial Neural Networks for cosmic gamma-ray propagation in the Universe}
\author[1]{K. K. Singh\corauthref{cor}},
\corauth[cor]{Corresponding author.}
\ead{kksastro@barc.gov.in}
	\author[1]{V. K. Dhar}, \author[2]{P. J. Meintjes}
\address[1]{Astrophysical Sciences Division, Bhabha Atomic Research Centre, Mumbai- 400 085, India} 
\address[2]{Physics Department, University of the Free State, Bloemfontein - 9300, South Africa}
\begin{abstract}
We explore the potential of an artificial neural network (ANN) based method 
intelligence to probe the propagation of cosmic $\gamma$-ray photons in the 
extragalactic Universe. The journey of $\gamma$-rays emitted from a distant 
source like blazar to the observer at the Earth is impeded by the absorption 
through the interaction with the extragalactic background light (EBL), leading 
to an electron-positron pair production. This process dominates for gamma ray photons 
with energy above 10 GeV propagating over the cosmological distances. The effect 
of $\gamma$-ray attenuation is characterized by a physical quantity called 
\emph{optical depth}, which strongly depends on the $\gamma$-ray photon energy, 
redshift of the source, and density of the EBL photons. We estimate the optical depth 
values for $\gamma$-ray energies above 10 GeV emitted from the sources at redshifts 
in the range 0.01 to 1 using three different and most promising EBL models. 
These optical depth estimates are randomly divided into two data sets for training 
and testing of the ANN using energy, redshift as inputs and optical depth as output. 
The optimization of ANN-performance for each EBL model employs standard back-propagation (BP) 
and radial-basis function (RBF) algorithms. The performance of the ANN model using the RBF is found 
to be superior to the BP method. In particular, the RBF-ANN with 40 neurons in the hidden layer 
corresponding to the EBL model proposed by Finke et al. (2010) shows the best performance for 
the propagation of $\gamma$-rays in the Universe.
\end{abstract}
\end{frontmatter}
\section{Introduction}  
In the last two decades, the machine learning approaches using artificial 
neural networks (ANNs) have received a significant attention for solving 
a variety of problems in astrophysics and cosmology [1,2,3,4,5,6,7]. 
Machine learning methods refer to the group of algorithms which can solve new problems on 
the basis of learning from the existing examples. They are generally divided into two main 
categories namely the supervised and unsupervised learning. In supervised learning, the algorithms 
learn from a correctly labeled data through the mapping of a set of features in the input data set to 
the output variable and construct a model from the input data [8]. The unsupervised algorithms learn 
from the complex relationships existing in the unlabeled data set comprising measured features and 
have several external free parameters [9]. The ANNs are introduced as a non-linear mathematical 
technique for data prediction on the basis of input training data. The ANN can solve problems which 
are based on non-linear correlations and pattern recognition. Some of the practically solvable 
problems using the ANN include object localization, forecasting, approximation, data compression, 
clustering, astrophysical classification, adaptive optics in the astronomical observations, parameter extraction 
for the 21cm Global signal, search for gravitational wave signals, modelling the dusty Universe etc. 
A hybrid model based on ANN has been proposed to predict spectral energy distributions of galaxies at wavelengths 
ranging from the optical to the radio [10]. Deep learning models, which make use of long interconnected layers of 
ANN with powerful learning algorithms, have also emerged as the state of the art techniques to achieve outstanding results in different 
fields from computer vision, image recognition to the astrophysics and fundamental research [11,12,13,14,15,16]. 

\par
The Universe comprises very powerful celestial $\gamma$-ray sources which emit a strong beam of photons 
with energy above few MeV. The $\gamma$ ray photons in the MeV-GeV energy range are characterized as 
high energy (HE) and those with energy above 100 GeV are referred to as very high energy (VHE) in modern 
astronomy [17]. The observations of celestial HE and VHE sources using the space and ground-based telescopes 
strongly depend on the nature of $\gamma$ ray photon propagation over cosmological distances in the Universe. 
The propagation of $\gamma$ ray photons emitted from the sources at cosmological redshifts is strongly affected 
by the presence of the extragalactic background light (EBL) [18,19] and the extragalactic magnetic field [20]. 
Due to the effect of EBL, the observed $\gamma$-ray spectra of the extragalactic sources differ significantly from 
the intrinsic spectra emitted at the source. This is attributed to the absorption of $\gamma$ ray photons by the 
low energy EBL photons via $\gamma-\gamma$ pair production [21,22]. However, oscillations between $\gamma$ ray photons 
and axion-like particles due to the extragalactic magnetic field lead to a reduction in the EBL absorption [20]. 
Therefore, the observed $\gamma$-ray spectra are the convolution of the photon-photon interactions and 
photon- axion-like particles oscillations. Both processes exhibit uncertainty due to a lack of precise measurement of the EBL 
spectral energy distribution, strength of the extragalactic magnetic field and existence of hypothetical axions beyond 
the standard model [23]. The EBL plays a dominant role in the propagation of $\gamma$ ray photons and limits the transparency 
of the Universe beyond a certain redshift [19]. It also introduces specific spectral signatures in the observed VHE spectra of 
the extragalactic sources. Therefore, the interpretation of $\gamma$-ray propagation in the extragalactic space is crucial to 
reconstruct the intrinsic VHE spectra of the sources which are indicative of the physical processes involved in the non-thermal emission. 
The propagation effects of the $\gamma$ ray photons also provide a way to indirectly probe the EBL which is an important cosmological 
quantity after the cosmic microwave background radiation  and characterizes the star formation and evolution of galaxies in 
the Universe [19,24,25]. Direct measurements of the EBL are challenged by the strong foreground contamination due to the zodiacal 
light and diffuse Galactic light [18,26]. 
\par
In this paper, we demonstrate that the ANN based module can be used as a tool to probe the propagation 
of $\gamma$ ray photons in the Universe. We use the opacity of the Universe caused by the EBL to $\gamma$-rays 
as a function of the observed photon energy ($E$) and source redshift ($z$) corresponding to three 
different popular EBL models for optimizing the performance of the ANN-module. We begin with a brief 
introduction to the EBL models in Section 2. In Section 3, we discuss the theory of $\gamma$-ray propagation 
over cosmological redshifts. The application of ANN in the $\gamma$-ray propagation is described in Section 4. 
Finally, we discuss and conclude the important findings of the study in Section 5.

\section{Extragalactic Background Light}
The EBL is the integrated diffuse radiation emitted from the stars and active galaxies through thermal and non-thermal 
processes and reprocessed by the interstellar dust over the history of the Universe [18]. Its spectral energy distribution 
ranges from the ultraviolet-optical (UV-Opt) to far-infrared (FIR) and exhibits two distinct humps. The first hump peaking 
at $\sim 1$ eV is characterized as the UV-Opt emission directly from the stars and the second one peaking at $\sim 0.01$ eV is 
ascribed as the IR emission due to the absorption and reprocessing of the stellar light by the dust present in the 
inter-galactic space. Therefore, the EBL spectrum carries important information about the star formation history 
and evolution of galaxies in the Universe. The density of the EBL photons is 10$^{-6}$ - 10$^4$)~ph~cm$^{-3}$~eV$^{-1}$ and 
their wavelength is redshifted due to the expansion of the Universe. The integrated EBL intensity represents a dominant 
component of the total energy budget of the Universe and the EBL contribution to the total energy density is $\sim 15\%$ of 
that due to the cosmic microwave background radiation [27]. 
The spectrum of the EBL at the present epoch is not completely known due to the uncertainties in the relevant astrophysical processes 
like stellar formation, evolution, reprocessing of light by cosmic dust and other gravitational and nuclear processes contributing to 
the EBL. However, several models for the spectral energy distribution of the EBL over a wide wavelength range have been developed using 
different empirical and computational approaches like forward evolution, backward evolution and semi-analytical [19,28]. A brief description of 
the three widely accepted and used EBL models is given below.

\subsection{Franceschini et al. (2008): Model-A} 
The EBL model proposed by Franceschini et al. (2008) is a backward evolution model based on the extrapolation of the low redshift 
galaxy data to higher redshifts under certain assumptions on the evolution of the galaxies in the Universe [29]. In this model, the 
EBL photon density and its evolution as a function of redshift are estimated from the fitting and interpolation of the multi-wavelength 
data available at UV-optical and IR wavelengths from various ground and space-based astronomical observations. Additional constraints on 
the spectral energy distribution of the EBL are obtained from the dedicated direct measurements at known wavelengths. Recently, the model 
has been reviewed on the basis of new optical and IR deep survey observations [30]. However, no significant modification on the EBL photon 
density is obtained at UV-optical and IR wavelengths. We have used the original EBL determinations reported in [29] in the present study 
and refer to this as \emph{Model-A}.

\subsection{Finke et al. (2010): Model-B}
In this model, the EBL intensity at UV-optical and IR wavelengths is estimated by assuming the main and post-main sequence stars as blackbodies 
and reprocessing of starlight by cosmic dust respectively [31]. This model is not based on the galaxy data and the evolution of galaxies 
over cosmological redshifts is inferred from the observed star formation rate density and initial mass function. The reprocessing of starlight
absorbed by dust is modelled as IR emissions by three blackbodies representing warm (large), hot (small) dust grains and polycyclic aromatic
hydrocarbons. The spectral energy distribution of the EBL estimated in [31] is close to the lower limits from the galaxy counts. 
This model provides the most accurate estimates for the density of the EBL photons at low redshifts and short wavelengths. We refer to 
this EBL spectral energy distribution as \emph{Model-B} in the present study.

\subsection{Dom\'inguez et al. (2011): Model-C}
Dom\'inguez et al. (2011) have derived the evolving spectral energy distribution of the EBL on the basis of observations only [32]. 
The novel method employed to estimate the EBL spectrum uses the observed evolution of the rest-frame galaxy luminosity function in K-band 
up to redshift $z~\sim$ 4 and changing fractions of the quiescent galaxies to fit the observed multi-wavelength templates. The evolving 
contribution to the bolometric EBL from UV to IR due to different populations of galaxies is calculated from the extrapolation of the galaxy 
spectral energy distribution type fractions. The modelling of EBL in [32] has two major shortcomings namely extrapolation of the galaxy 
fractions at higher redshifts and colour dependent selection effects. The EBL spectrum obtained by this model is consistent with the majority 
of the data from galaxy counts. We define this as \emph{Model-C} in the present study.
\par
\begin{figure}
\begin{center}
\includegraphics[width=0.70\textwidth,angle=-90]{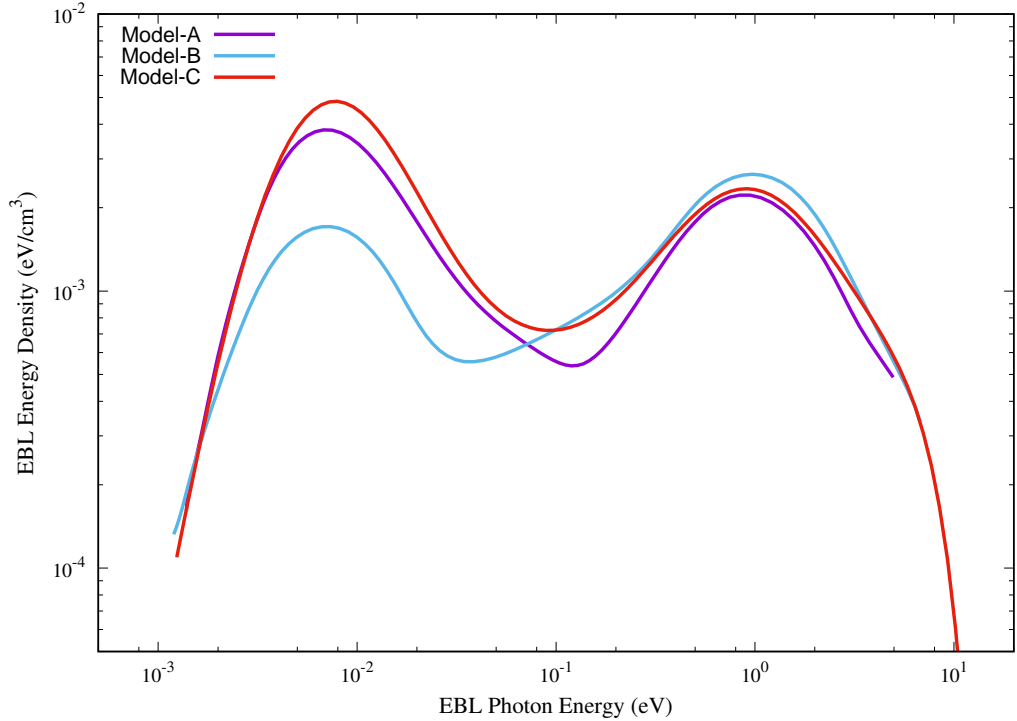}
	\caption{Spectral energy distributions of the EBL proposed by three different models: \emph{Model-A} [29], 
	\emph{Model-B} [31], and \emph{Model-C} [32].}
\label{fig:ebl}
\end{center}
\end{figure}

The energy density of EBL photons predicted by the above models is shown in Figure \ref{fig:ebl}. These models are generally 
used in the study of the opacity of the Universe to VHE $\gamma$ ray photons. Therefore, we have employed these three models in 
the present work to explore the potential of ANN in probing the $\gamma$-ray propagation in the Universe.

\section{$\gamma$-ray Propagation}
The propagation of $\gamma$ ray photons over cosmological distances is hindered due to the presence of EBL via $\gamma - \gamma$ absorption. 
The $\gamma$ ray photons emitted from a distant cosmic source and propagating in the intergalactic space can be absorbed by the EBL photons 
through the Breit-Wheeler process which is expressed as [33]
\begin{equation}
	\gamma~~+~~\gamma_{EBL}~~\rightarrow~~e^-~~+~~e^+
\end{equation}	
This process is kinematically allowed when the following threshold condition is satisfied [34]
\begin{equation}\label{eqn:ppthres}
	E~\epsilon~=~\frac{2 m_e^2 c^4}{(1 + z)^2 (1-\rm cos \rm \theta)}
\end{equation}
where $E$ and $\epsilon$ are the energies of $\gamma$-ray and EBL photons measured at the present epoch respectively, $\theta$ is the 
angle between the directions of the motion of two photons and $m_e c^2$ is the rest mass energy of the electron. The $(1+z)^2$ factor 
takes into account the change in energy of photons due to the cosmic expansion of the Universe. For an isotropic distribution of the 
EBL photons, the minimum energy of the background photons for the maximum $\gamma - \gamma$ absorption is 
given by [34]
\begin{equation}
	\epsilon~\approx \left(\frac{900}{(1+z)^2 E(GeV)}\right)~\rm eV
\end{equation}	
This suggests that for VHE $\gamma$-ray photons with energy in the range 100 GeV $\le E \le$ 20 TeV emitted from the sources at  
redshift $z \sim$ 1, the energy of the background photons for the absorption via pair production should be 2.2 eV $\le \epsilon \le$ 0.011 eV. 
Therefore, the EBL photons play a dominant role in the absorption of VHE $\gamma$-ray photons propagating over cosmological distances 
($z >$ 0) in the Universe. This process leads to an energy ($E$) and redshift ($z$) dependent depletion of the VHE $\gamma$-ray flux from a 
distant cosmic source and therefore limits the transparency of the Universe to $\gamma$-rays. The survival probability (i.e probability of 
escaping from the EBL absorption) of VHE $\gamma$-ray photons is estimated as 
\begin{equation}
	P_{\gamma \rightarrow \gamma}~=~e^{-\tau (E, z)}
\end{equation}	
where $\tau$ is referred to as the \emph{optical-depth}, which characterizes the opacity of the Universe and strongly depends on 
$E$ and $z$ for a given spectral energy distribution of the EBL photons. By definition, $\tau (E,z) = 1$, is termed as 
``$\gamma$-ray horizon'' and $\tau (E, z)<$ 1 or $\tau (E, z) >$ 1 corresponds to transparent or opaque Universe respectively. 
The optical depth $\tau (E,z)$ for a given EBL model is computed from the following three fold integral [25,34]
\begin{equation}\label{eqn:tau}
	\tau (E,z_s)~=~\int_0^{z_s} \frac{dl}{dz}~dz \int_0^{\pi} \left(\frac{1 - \rm{cos \theta}}{2}\right)\rm{sin \theta} d\theta 
		     \int_{\epsilon_{min}}^\infty n(\epsilon,z) \sigma_{\gamma\gamma}(E,\epsilon,\theta) d\epsilon
\end{equation}	
where $z_s$ is redshift of the $\gamma$-ray source, $\left(\frac{dl}{dz}\right)$ is the line element for the distance travelled by a VHE photon, 
n$(\epsilon, z)$ is the number density of the EBL photons per unit energy, and $\sigma_{\gamma\gamma}(E,\epsilon,\theta)$ is the total scattering 
cross section for pair production. The lower limit of the integral over the energy of EBL photons ($\epsilon_{min}$) is given by the threshold condition 
for pair production from Equation \ref{eqn:ppthres}. We have used the following set of parameters in the $\Lambda$CDM cosmology
\begin{equation}
	\frac{dl}{dz}~=~\frac{c}{H_{0}}\frac{1}{(1+z)\sqrt{\Omega_{\Lambda}+\Omega_{m}(1+z)^{3}}}
\end{equation}
with $\Omega_m =$ 0.30, $\Omega_\Lambda =$ 0.70 and H$_0 =$ 70~km~s$^{-1}$~Mpc$^{-1}$ in the present work. 

\begin{figure}
\begin{center}
\includegraphics[width=1.0\textwidth]{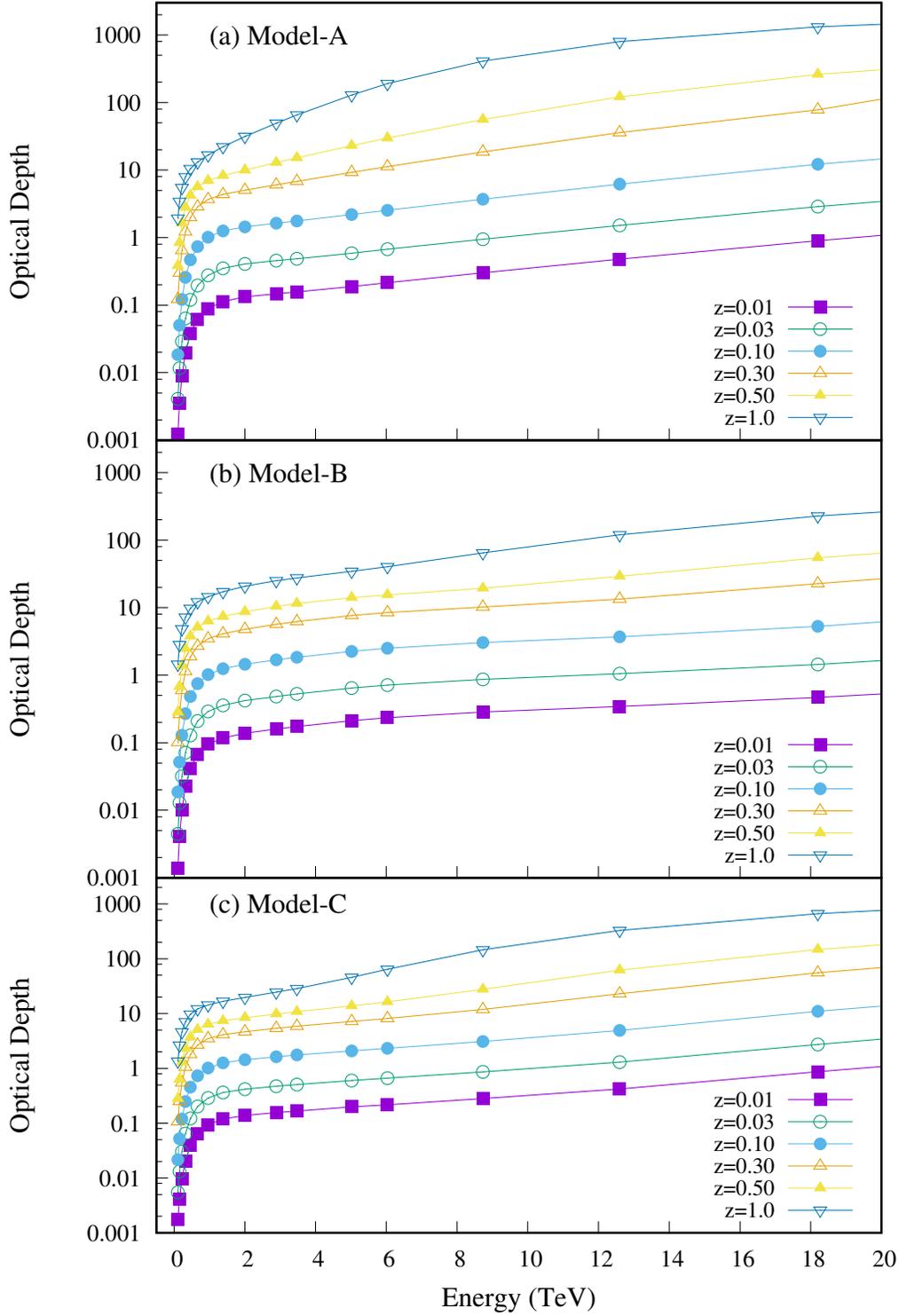}
	\caption{Data set employed for training the ANN using three different EBL models. A similar data set is used for testing of the ANN.}
\label{fig:train}
\end{center}
\end{figure}

\section{Application of ANN to the gamma-ray propagation}
The ANN represents a simplified model of the human brain functioning and the nervous system. 
A human brain consists of more than 10$^{10}$ neurons connected to each other with a complex 
topology and a simple ANN is described as a network of three inter-connecting layers namely 
input layer, hidden layer and output layer [35]. Each layer comprises of different number of neurons as 
basic building blocks and the number of neurons in the hidden layer determines the width of the ANN [36,37]. 
An ANN having several hidden layers due to the complexity of the problem is referred to as deep ANN. 
Each neuron in the input layer (first layer) is connected to one or more neurons in the hidden layer 
(intermediate layer). The individual connections between the neurons of the input and hidden layers are 
characterized by an appropriate weight and bias [38]. Each neuron in the hidden layer derives an \emph{activation} 
through the linear combination of the input neurons from the input layer and the weights and biases associated 
with the respective connections. These activations are passed through linear or non-linear transformations to 
the output layer which performs a simple sum of all inputs from the hidden layer and predicts an output. 
This is referred to as the \emph{feed-forward process} in the simple ANN model [39]. The width, deepness and 
activations of an ANN are decided in the beginning of the model, whereas weights and biases associated with the 
connections are free parameters and have to be optimized during the training process. To optimize the performance 
of an ANN model, a loss function or error function is defined. In the present work, we have used \emph{mean squared error} (MSE) 
to measure the performance of the ANN model for the $\gamma$-ray propagation. It is defined as 
\begin{equation}\label{eqn:mse}
	\rm MSE~=~\frac{1}{N}\Sigma_{i=1}^N \left(Y_d^i - Y_p^i \right)^2
\end{equation}	
where $N$ is the total number of input samples, $Y_d^i$ and $Y_p^i$ are the desired (true) and ANN predicted 
outputs respectively for the $i^{th}$ sample. The mean squared error loss function gives outliers a larger weightage 
and needs to be minimized during the training for determining the free parameters like weights and biases in an 
excellent performing ANN. Different types of supervised and unsupervised algorithms are used to simulate the 
learning process of the ANN. In this study, we have employed two algorithms namely the standard \emph{Back-propagation} 
(BP) and \emph{Radial basis function} (RBF) to construct the neural network for probing the $\gamma$-ray propagation 
in the Universe. We have used an in-house developed software package BIKAS (Barc Iit KAnpur Simulator) for a major 
portion of the present work. Dhar et al. (2010) have reported a comparative study of a few popular ANN algorithms on 
the benchmark and function approximation problems using BIKAS package [40]. A brief and qualitative description of the BP 
and RBF algorithms used in the present study is given below.

\subsection{Back-propagation} 
The back-propagation (BP) is a supervised learning algorithm which involves the gradient-descent paradigm to minimize 
the mean squared error [41]. In this algorithm, the weights are randomly initialized to small values to calculate the ANN output ($Y_p^i$) and 
compare this with the desired output ($Y_d^i$) value over every $i^{th}$ training input. If the values of $Y_p^i$ and $Y_d^i$ are 
not similar, the error or difference between these two values is computed. The errors are propagated backward from the output layer 
to the input layer after every epoch and the weights are adjusted according to the \emph{Delta Rule} for minimizing the error value. 
This employs the modification of weights along the most direct path in the weight space through the slope or gradient of the 
error function. The changes in weights towards the error minimization are proportional to the negative gradient of the error 
function [42]. This process repeatedly adjusts the weights in the ANN until the mean squared error at the network ouput is minimized. 
The gradient-descent based BP is a popular method for training of the ANN due to its simplicity and operational ease. However, 
the selection of hidden layers is critical in the design of a BP-based neural network since there is no analytical approach 
for optimizing the deepness of an ANN. This is generally done through the proper training of the ANN on various permutations and 
selection of the minimum number of layers or neurons which can yield the lowest value of the mean squared error. In addition to this, the training of 
the BP-based ANN is quite time consuming and exhibits the local minima entrapment problem.

\subsection{Radial Basis Function}
The radial basis function (RBF) provides an efficient and attractive alternative to BP method for the training of 
the ANN [43,44]. The RBF-based ANN has a capability of universal function approximation and regularization. Unlike BP, RBF 
approach constructs approximation models assuming linear terms of the parameters or weights for any continuous function [45]. 
The training of the RBF-based ANN is faster than that of the BP since it is constrained to have just two layers of weigths which 
are sequentially fixed. The whole structure of the network is fixed only through the determination of the hidden layer and the weights 
associated with the connections between the hidden layer and ouput layer. The weights associated with the connections between the 
input layer and hidden layer are decided after fixing the hidden layer. The weights of the hidden layer neurons are assigned in 
such a way that the network produces the maximum output for inputs equal to their weights and are not initialized randomly unlike BP. 
The RBF algorithm derives its working methodology from the famous \emph{Cover's Theorem}, which states that \emph{a linearly non-separable 
data set in a low dimensional space can be transformed into a linearly separable data set in a high dimensional space through some 
non-linear transformation}. The most commonly used activation function in the RBF-based ANN is a Gaussian function. The RBF algorithms 
do not exhibit any of the BP training anomalies like the problem of local minima.
\par
The main objective of this work is to explore the potential of the ANN for probing the propagation of $\gamma$-ray photons 
over cosmological distances in the Universe. As discussed in Section 3, the $\gamma$-ray propagation is characterized by 
the optical depth ($\tau$) which is a function of the redshift ($z$) and energy of the photon ($E$) for a given EBL model. 
We have estimated $\tau$ values for the $\gamma$-ray photons in the energy range $E =$ 10 GeV - 20 TeV at redshifts $z =$ 0.01, 
0.03, 0.10, 0.30, 0.50, 1.0 corresponding to the three different and most promising EBL models designated as Model-A, Model-B, 
and Model-C. We randomly split this data set into \emph{Training} and \emph{Testing} samples for optimization of the ANN performance 
using BP and RBF algorithms. The data samples used for training of the ANN for three EBL models are depicted in Figure \ref{fig:train}. 
A similar data set is generated for the purpose of testing the ANN.

\begin{table}
\caption{Summary of the mean squared error for optimization of the number of neurons using the BP algorithm for Model-A.}
\vspace{0.5cm}
\begin{center}
\begin{tabular}{lccc}
\hline
Number of Neurons	&MSE			&$\sigma$(Training)	     &$\sigma$(Testing)\\
\hline
5  			&3560.81               &512.62     			&788.32\\   
10  			&681.09                &213.75     			&311.81\\ 
15			&144.46                &97.35     			&128.52\\
20      		&47.49                 &40.42     			&78.51\\
25      		&17.49                 &26.21     			&42.96\\    
30			&8.02                  &20.51     			&31.91\\ 
{\bf 35}		&{\bf 2.14}	       &{\bf 13.21} 			&{\bf 18.36}\\
40			&0.34                  &7.82     			&46.53\\
\hline
\end{tabular}
\end{center}
\label{tab:bp-Frances}
\end{table}

\begin{table}
\caption{Same as Table \ref{tab:bp-Frances} for Model-B.}
\vspace{0.5cm}
\begin{center}
\begin{tabular}{lccc}
\hline
Number of Neurons	&MSE			&$\sigma$(Training)	     &$\sigma$(Testing)\\
\hline
5  			&212.33               &79.22     			&92.15\\   
10  			&63.026               &54.99     			&79.43\\ 
15			&14.024               &36.37     			&66.79\\
20      		&3.38                 &27.96     			&42.12\\
25      		&0.63                 &22.10     			&31.47\\    
30			&0.28                 &18.49     			&26.51\\ 
{\bf 35}		&{\bf 0.09}	      &{\bf 8.92} 			&{\bf 10.78}\\
40			&0.02                 &6.41     			&12.47\\
\hline
\end{tabular}
\end{center}
\label{tab:bp-Finke}
\end{table}

\begin{table}
\caption{Same as Table \ref{tab:bp-Frances} for Model-C.}
\vspace{0.5cm}
\begin{center}
\begin{tabular}{lccc}
\hline
Number of Neurons	&MSE			&$\sigma$(Training)	     &$\sigma$(Testing)\\
\hline
5  			&1366.95                &460.51     			&635.92\\   
10  			&164.13                 &212.82     			&501.32\\ 
15			&28.20                  &102.37     			&332.46\\
20      		&8.19                   &61.92     			&212.31\\
25      		&1.22                   &41.36     			&101.32\\    
30			&0.32                   &27.21     			&47.21\\ 
35		        &0.08	                &19.22 			        &27.92\\
{\bf 40}		&{\bf 0.132}            &{\bf 7.21}     		&{\bf 13.29}\\
\hline
\end{tabular}
\end{center}
\label{tab:bp-Doming}
\end{table}

\begin{table}
\caption{Summary of the mean squared error for optimization of the number of neurons using the RBF algorithm for Model-A.}
\vspace{0.5cm}
\begin{center}
\begin{tabular}{lccc}
\hline
Number of Neurons	&MSE			&$\sigma$(Training)	     &$\sigma$(Testing)\\
\hline
5  			&113.56               &368.51     			&507.48\\   
10  			&24.19                &55.74     			&80.30\\ 
15			&1.16                 &23.39     			&42.13\\
20      		&0.76                 &17.54     			&31.12\\
25      		&0.46                 &9.15     			&30.36\\    
30			&0.24                 &7.92     			&11.21\\ 
{\bf 35}		&{\bf 0.22}           &{\bf 5.21} 			&{\bf 11.81}\\
40		        &0.18                  &3.41     			&15.42\\
\hline
\end{tabular}
\end{center}
\label{tab:rbf-Frances}
\end{table}

\begin{table}
\caption{Same as Table \ref{tab:rbf-Frances} for Model-B.}
\vspace{0.5cm}
\begin{center}
\begin{tabular}{lccc}
\hline
Number of Neurons	&MSE			&$\sigma$(Training)	     &$\sigma$(Testing)\\
\hline
5  			&122.32                &66.61     			&78.36\\   
10  			&37.30                 &38.42     			&67.11\\ 
15			&9.02                  &24.62     			&42.54\\
20      		&1.27                  &19.96     			&32.93\\
25      		&0.49                  &16.39     			&26.43\\    
30			&0.22                  &12.23     			&21.51\\ 
35			&0.14	               &9.71 				&15.02\\
{\bf 40}		&{\bf 0.07}            &{\bf 7.09}     			&{\bf 8.47}\\
45			&0.05                  &6.32     			&14.87\\
\hline
\end{tabular}
\end{center}
\label{tab:rbf-Finke}
\end{table}

\begin{table}
\caption{Same as Table \ref{tab:rbf-Frances} for Model-C.}
\vspace{0.5cm}
\begin{center}
\begin{tabular}{lccc}
\hline
Number of Neurons	&MSE			&$\sigma$(Training)	     &$\sigma$(Testing)\\
\hline
5  			&79.26                 &36.18     			&84.97\\   
10  			&29.81                 &21.16     			&67.14\\ 
15			&7.36                  &17.41     			&34.21\\
20      		&2.31                  &13.51     			&21.14\\
25      		&0.98                  &10.94     			&15.92\\    
{\bf 30}		&{\bf 0.13}            &{\bf 7.69}     			&{\bf 9.06}\\ 
35			&0.02	               &3.24 				&11.18\\
\hline
\end{tabular}
\end{center}
\label{tab:rbf-Doming}
\end{table}

\subsection{Training and Testing of the ANN}
The most important requirement for the training of an ANN module is that a well trained ANN should work satisfactorily on a new input data set 
(testing) which was not used in the training process. Therefore, a high quality training data is crucial for the optimization of an ANN module 
for further applications. In this work, the training data set consist of $E$ and $z$ as inputs and $\tau$ as the desired ouput for three different 
EBL models (Figure \ref{fig:train}). The first step in the training stage involves determining the optimized number of neurons to be employed in 
the ANN module. We have followed a trial and error method for optimizing the number of neurons during the training process [46]. 
The details of the optimization process for the number of neurons by minimizing the mean squared error using the BP and RBF algorithms are given in 
Table (\ref{tab:bp-Frances}-\ref{tab:bp-Doming}) and Table (\ref{tab:rbf-Frances}-\ref{tab:rbf-Doming}) respectively for three different EBL models 
with 12000 iterations in each case. 

\begin{figure}
\begin{center}
\includegraphics[width=0.70\textwidth,angle=-90]{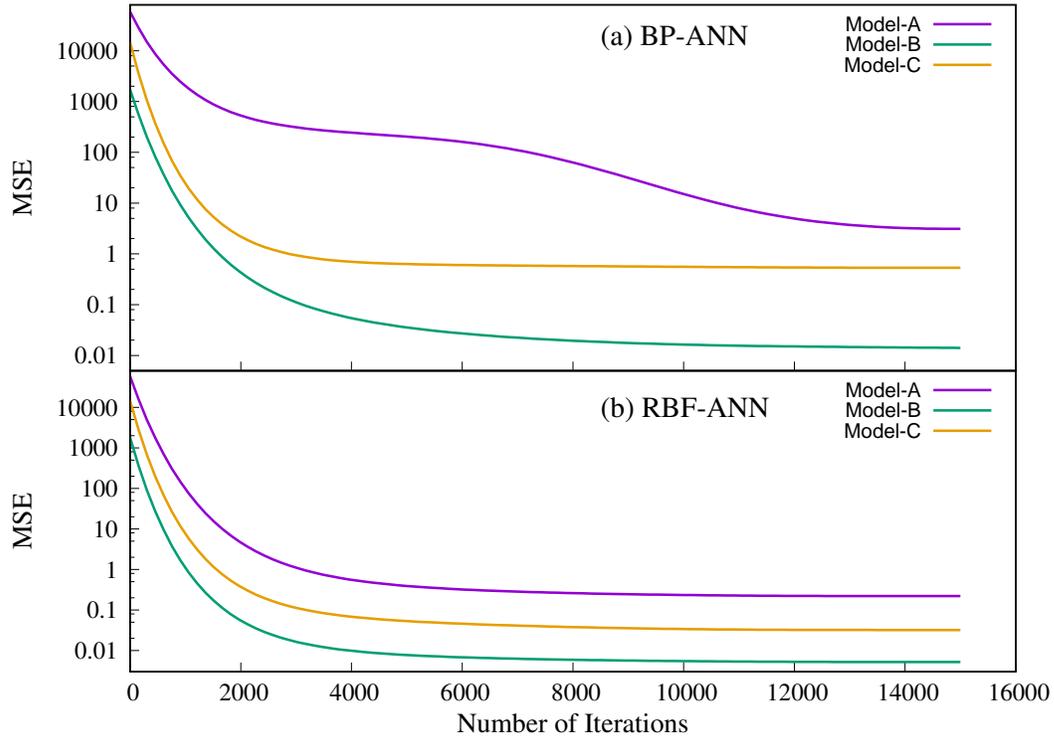}
\caption{Mean Squared Error during the training of ANN using two different algorithms (a) Back-propagation (BP) and 
	(b) Radial basis function (RBF), corresponding to three EBL models: \emph{Model-A}, \emph{Model-B}, and \emph{Model-C}.}
\label{fig:mse}
\end{center}
\end{figure}

\begin{figure}
\begin{center}
\includegraphics[width=0.70\textwidth,angle=-90]{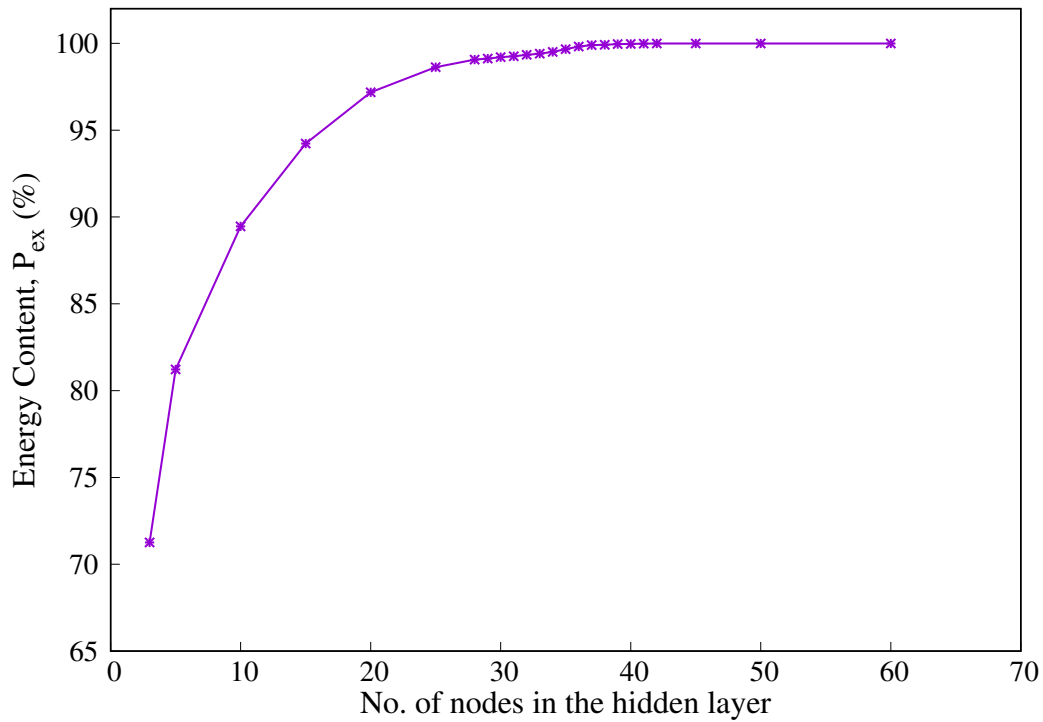}
\caption{Percentage of energy explained as a function of number of neurons in the hidden layer for the RBF-ANN corresponding to the 
                 \emph{Model-B} of the EBL.}
\label{fig:svd}
\end{center}
\end{figure}

We have also computed the standard deviation ($\sigma$) of the errors between expected output and ANN predicted value 
for the training as well as testing data sets. It is important to mention here that the BP is a standard algorithm for minimization 
of the mean squared error in neural networks. Therefore, we have compared the performance of these two algorithms in the present work. 
The combinations of the optimized number of neurons corresponding to the minimum mean squared error values and a close matching between 
the $\sigma$-values for the training and testing data sets are shown as boldfaces in Tables (\ref{tab:bp-Frances}-\ref{tab:rbf-Doming}). 
We note that a close matching between the $\sigma$-values occurs for 35 neurons (in one hidden layer) for Model-A and Model-B and for 40 neurons 
for Model-C in the case of BP method in the present study (Table \ref{tab:bp-Frances}-\ref{tab:bp-Doming}). Whereas, in the case of RBF algorithm, 
the corresponding number of neurons are 35, 40 and 30 for Model-A, Model-B and Model-C respectively (Table \ref{tab:rbf-Frances}-\ref{tab:rbf-Doming}). 
These are referred to as optimum neurons and are further used to optimize the number of iterations, which is relatively an easier task. The resulting 
mean squared error as a function of the number of iterations for the BP and RBF algorithms is shown in Figure \ref{fig:mse} (a) and (b) respectively. 
It is evident from the figure that the least mean squared error is observed for the Model-B after $\sim$ 10000 iterations for BP as well as RBF methods. 
A relatively worst performance is obtained for the Model-A with both minimizing methods. The minimum values of the mean squared error associated with 
the Model-B are 0.0963 and 0.0780 for the BP and RBF methods respectively. This implies that the performance of the RBF algorithm is better than the 
conventional BP method after about 12000 iterations for the application to $\gamma$-ray propagation problem. Since the number of neurons in the hidden 
layer is crucial in the design of an ANN, we have further employed the Singular Value Decomposition (SVD) method to validate the optimum neurons [6,47,48]. 
We consider Model-B for the SVD method because the ANN based on this model shows the best performance as discussed above. We deliberately 
use larger number of neurons  (60 in the present case) in the hidden layer to generate the weight matrix ($F$) from the output of each node 
before subjecting them to the non-linear transformations. With a total of 102 training patterns and one hidden layer of 60 neurons, the order of 
matrix $F$ will be 102$\times$60. The singular value decomposition of matrix $F$ is given by [49]
\begin{equation}
	F~=~U~S~V^T
\end{equation}
where $U$ and $V$ are the orthogonal matrices, and $S$ is a digonal matrix with 102 rows and 60 columns. The matrix $S$ contains singular 
values of the weight matrix $F$ on its diagonal. The dominance of the significant singular values of $F$ is determined by the percentage of 
energy explained ($P_{ex}$) which is defined as [49]
\begin{equation}
	P_{ex}~=~\frac{\Sigma_{i=1}^q S_i^2}{\Sigma_{i=1}^p S_i^2}~\times~100
\end{equation}
where $S_1$,$S_2$, $S_3$.....$S_p$ are the singular values of $F$ in the descending order. $p$ and $q$ represent total and dominant 
number of singular values respectively. $P_{ex}$ as a function of number of nodes in the hidden layer is shown in Figure \ref{fig:svd} for 
the RBF based ANN using Model-B. It is obvious from the figure that 99.99$\%$ energy is retained with 37 neurons as compared to 40 neurons 
estimated from the mean squared error minimization (Table \ref{tab:rbf-Finke}). Therefore, 37-40 neurons in principle can be considered as 
optimal neurons in the present study.

\par
\begin{figure}
\begin{center}
\includegraphics[width=1.0\textwidth]{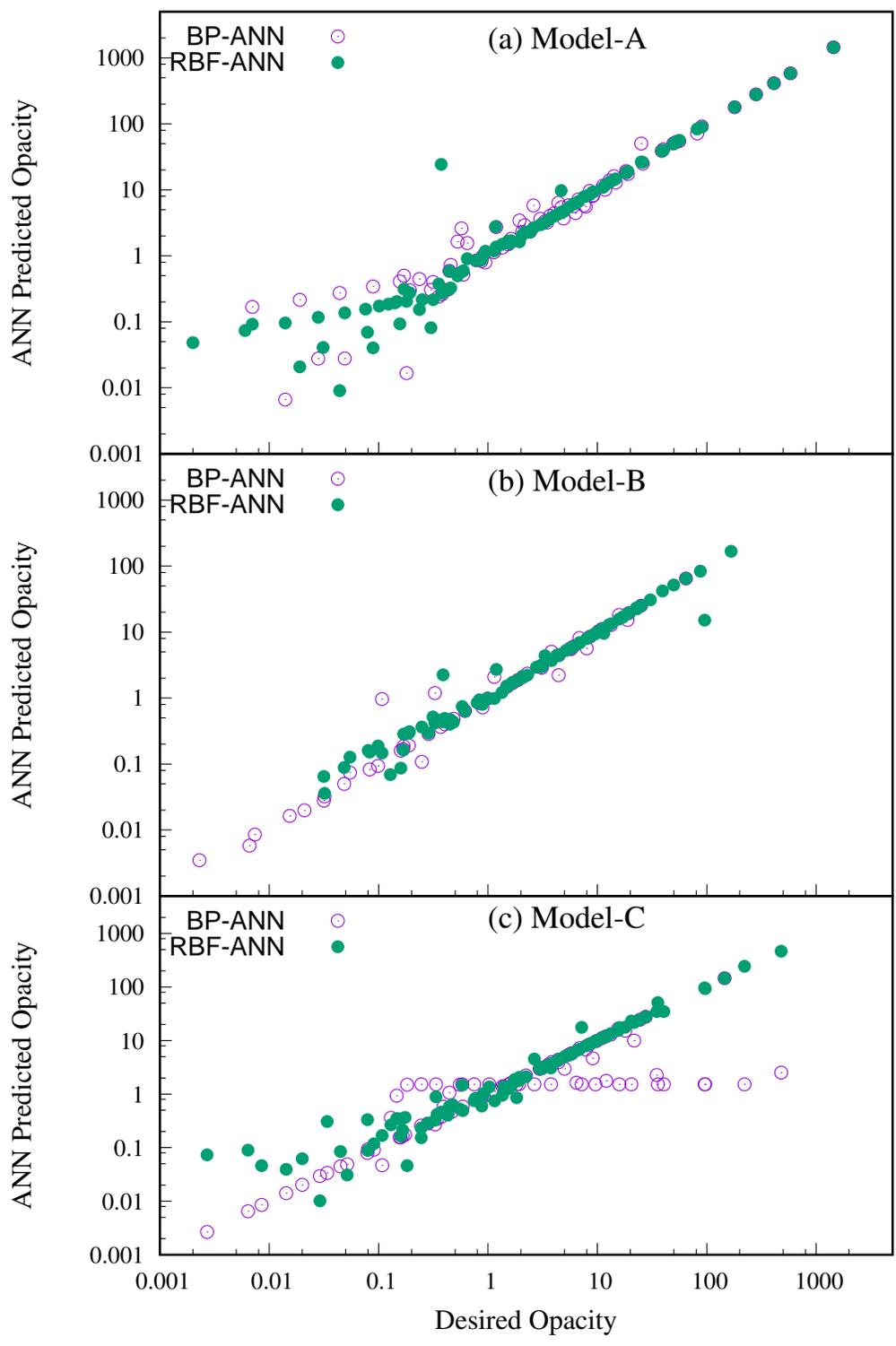}
\caption{Comparison of the optical depth values expected from three different EBL models with the corresponding values predicted by 
	the BP and RBF-based ANN.}
\label{fig:tau}
\end{center}
\end{figure}

In order to further monitor the performance of the ANN for the propagation of $\gamma$-ray photons, we randomly select different 
combinations of $E$ and $z$ to predict the optical depth values by the above trained ANN corresponding to three EBL models. We also 
compute the corresponding $\tau$ values using Equation \ref{eqn:tau} for Models-A, B, and C. A comparison of the ANN predicted 
optical depth values with the desired values (from Equation \ref{eqn:tau}) for three EBL models is depicted in 
Figure \ref{fig:tau}(a-c). This forms independent data sets (i.e. unseen during training and testing procedures) to measure the robustness 
of the BP and RBF-based ANN designed for the propagation of $\gamma$-rays in the present work. We observe from the scatter plot in 
Figure \ref{fig:tau}(a-c) that the optical depth values predicted by the BP and RBF-based ANN show a better linear correlation with 
those expected from a given EBL density for Model-B (Figure \ref{fig:tau}b), whereas a large deviation is found for Models-A 
and C (Figure \ref{fig:tau}a \& c). It is evident from Figure \ref{fig:mse} and Figure \ref{fig:tau} that the BP-algorithm does not 
converge in general and especially for Model-C (Figure \ref{fig:mse}a and Figure \ref{fig:tau}c). This can be attributed to the fact 
that the BP-algorithm is based on the gradient-descent method, which is known to suffer from the problems of getting stuck in the 
local minima and slow convergence. On the other hand, convergence of global minima is more probable in the second order methods like RBF 
and therefore these algorithms are inherently better suited for the problems with non-linearity. Also, the performance of all ANNs is worse 
at low opacity values. At lower opacity values, change in the mean squared error is small despite large percentage error. This can be the 
result of the mean squared error definition implemented in the present study. Since the mean squared error values are fed back to optimize 
the weights, the contribution of smaller mean squared error  values is less in modifying the weights and hence the training is restricted. 
To quantify the strength of the linear correlation,  we have also computed the 
Pearson coefficient using the ANN predicted and desired optical depth values for Model-B. We obtain Pearson coefficient values of 
$\sim$0.48 and $\sim$0.93 for the BP and RBF methods respectively. This indicates that the $\tau(E,z)$ values predicted by the RBF-based ANN 
show a strong and positive correlation with those expected from the Model-B. Therefore, the performance of the RBF-based ANN with 40 optimum 
neurons for the propagation of $\gamma$-rays over cosmological distances in the framework of Model-B is found to be better than the BP-based ANN 
with 35 neurons in the present work. The number of neurons is an important consideration in the design and application of ANN to any problem at hand. 
If the number of neurons employed in the ANN is less or more than the optimal neurons, the network is accordingly said to be under or over trained. 
This can result in over fitting or remembering the training data rather than its generalization [7]. 

\section{Discussions and Conclusions}
\begin{figure}
\begin{center}
\includegraphics[width=0.70\textwidth,angle=-90]{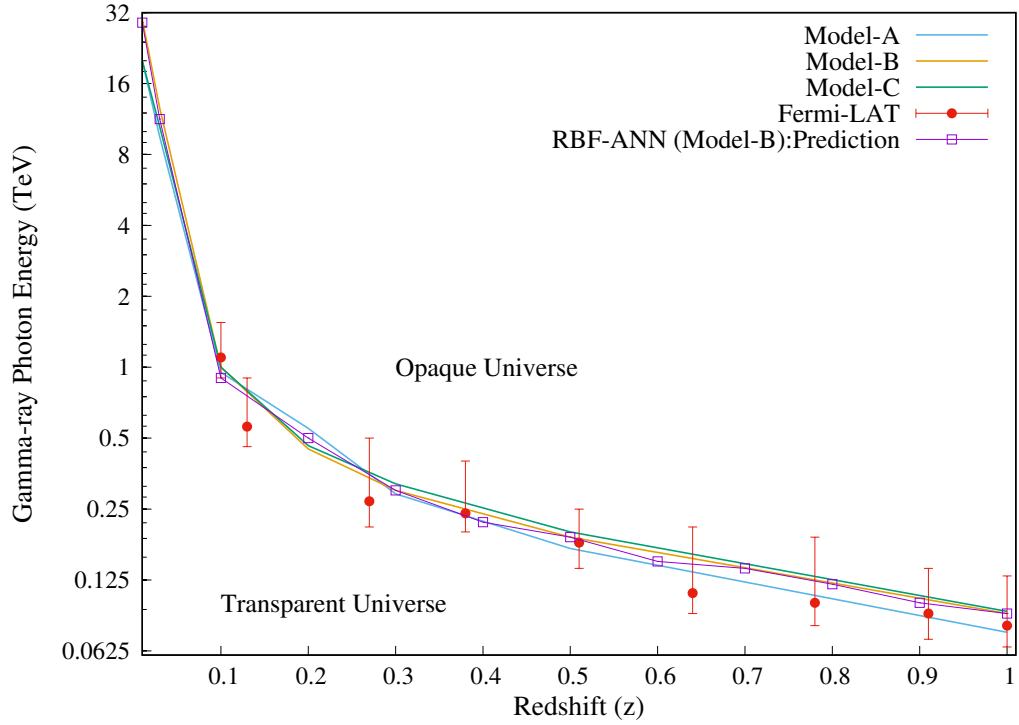}
	\caption{Comparison of the cosmic gamma-ray horizon predicted by the RBF-based ANN designed using Model-B [31] 
		and from the three different EBL models with the \emph{Fermi-LAT} observations [54].}
\label{fig:horizon}
\end{center}
\end{figure}

We have used optical depth estimates from the three different EBL models for a comparative study of the performance of the BP 
and RBF based ANN in the propagation of $\gamma$ ray photons over cosmological distances in the Universe. A properly designed ANN 
predicts an output corresponding to a set of inputs after a proper training by a large number of a priori known set of inputs and 
their desired outputs. The propagation of $\gamma$ ray photons with an energy over 10 GeV is strongly affected by the large uncertainties 
in the density of ultraviolet, optical and infrared components of the EBL photons. The present study suggests that an RBF-based ANN outperforms 
the conventional BP method for probing the opacity of the Universe to the $\gamma$-rays emitted from a source in the redshift 
range $z = 0.01 -1$. Among the three most promising EBL models considered in this study, we find that the overall performance of 
the RBF-based ANN module is in better agreement with the EBL model proposed by Finke et al. (2010). Recently, the optical depth 
estimates from a model-independent approach based on the GeV-TeV observations of blazars at various redshifts also suggest that the models 
proposed by Finke et al. (2010) and Dom\'inguez et al. (2011) give a relatively better description of the spectral energy distribution 
of the EBL than many other models in the literature [50]. The combined observations of blazars from the \emph{Fermi}-Large Area Telescope (LAT) 
and the ground-based Cherenkov telescopes have provided a unique tool to probe and constrain the density of EBL photons [51,52,53]. 
The role of the spectral energy distribution of EBL is extremely important in cosmology to probe the galaxy formation and evolution throughout 
the history of the Universe. The evolution of EBL has been recently reconstructed by using the \emph{Fermi}-LAT observations of more than 700 active 
galaxies in the redshift range  $z = 0.03 -3$ [54,55]. The star formation history determined from the \emph{Fermi}-LAT measurements is in good agreement 
with the independent measurements from galaxy surveys with a peak at $z \sim 2$. A comparison of the cosmic $\gamma$-ray horizon ($\tau (E, z) =1$) from 
the \emph{Fermi}-LAT observations up to $z =1 $ with the predictions from Models-A, B \& C and by the RBF-ANN corresponding to Model-B is 
shown in Figure \ref{fig:horizon}. The \emph{Fermi}-LAT measurements provide $\gamma$-ray horizon up to 1 TeV only due to its observational 
energy range of 0.1 GeV to less than 2 TeV. In this energy range, the cosmic $\gamma$-ray horizon predicted by the RBF-ANN corresponding to 
Model-B is broadly consistent with the \emph{Fermi}-LAT measurements. Future $\gamma$-ray observations with the Cherenkov Telescope Array (CTA) over 
a wide energy range will be extremely helpful in probing the transparency of the Universe to higher redshifts and subsequently the spectral energy 
distribution of the EBL photons. This will enable a better understanding of the $\gamma$-ray propagation problem in cosmology. 
\par
In summary, the present study suggests that the RBF-based ANN module having 40 neurons in the hidden layer can describe the propagation of 
$\gamma$-rays in the Universe in a better agreement with the EBL model proposed by Finke et al. (2010). 
To the best of our knowledge, the application of RBF or BP-based ANN methodology for exploring the $\gamma$-ray propagation over the cosmological 
distances is demonstrated for the first time in the present work. The RBF algorithm being a powerful method, is found to be suitable for recovering the 
cosmic optical depth to $\gamma$ ray photons due to interaction with the EBL photons. Apart from estimation of the optical depth for 
a given $\gamma$-ray energy and source-redshift, the RBF-based ANN designed in the present work can also be effectively employed to explore the compatibility of 
the various existing EBL models with the independent measurements from $\gamma$-ray observations. Out of three prominent EBL models used in this study, 
Finke et al. (2010) model yields a solid result, where RBF-based ANN reproduces the opacities over the full parameter space and also in good agreement with the 
\emph{Fermi}-LAT measurements. Therefore, the RBF-based methodology can be potentially used as an additional probe for different EBL models as it involves 
both the direct EBL measurements and the indirect opacities predicted by the $\gamma$-ray observation. Thus, the new knowledge contributed by the present 
study lies in the fact that the RBF-approach is most suitable to demonstrate the potential of ANN for probing the transparency of Universe to high energy 
$\gamma$-rays in general. Almeida et al. have successfully developed a new ANN approach in generating accurate predictions of the spectral energy distribution 
of the EBL photons using semi-analytical galaxy formation models [10]. This hybrid model can produce mock catalogues of galaxies for the forthcoming surveys 
for the extragalactic background radiation. 
\par
The performance of ANN in the present work can be further improved by including other physical processes related to the $\gamma$-ray propagation like photon- axion-like particles 
oscillations and cascading due to the presence of the extragalactic magnetic field. In the photon- axion-like particles coupling, VHE $\gamma$ ray photons in 
a beam can partially convert to axion-like particles during their journey from the source to the Earth in the presence of a coherent extragalactic magnetic 
field oriented in the direction transverse to the photon propagation [20]. The  axion-like particles travel unimpeded over cosmological distances due to the 
negligibly small scattering cross section like neutrinos and convert back to the original $\gamma$-ray photons in the Galactic (Milky Way) magnetic field 
before reaching the Earth [20]. This process, generally referred to as the \emph{photon- axion-like particles oscillation}, may enhance the transparency 
of the Universe to VHE $\gamma$ rays. Consequently, the effective value of $\tau (E,z)$  will be reduced and the photon survival probability increases. 
Some observations with the modern ground and space-based instruments indicate that the Universe is more transparent to $\gamma$-rays than expected from the 
EBL absorption [56,57,58]. However, no direct experimental evidence regarding the existence of the  axion-like particles has been obtained so far and only 
their parameter space (mass and coupling constant) is constrained at a given confidence level [59,60,61]. Also, the strength and coherence length of the 
extragalactic magnetic field are not clearly known so far and remain currently an active area of research.

\section*{Acknowledgement}
Authors thank the anonyomus reviewer for her/his valuable suggestions to improve the contents of the manuscript.

\end{document}